\documentclass[12pt]{article}
\title{\vspace{-2cm}
A comparison of four approaches to the calculation of conservation laws}

\author{Thomas Wolf\\Department of Mathematics, Brock University\\
       St. Catharines, ON   L2S 3A1, Canada}

\begin{document}
\maketitle

\begin{abstract}
The paper compares computational aspects of four approaches to compute
conservation laws of single differential equations (DEs) or systems of
them, ODEs and PDEs. The only restriction, required by two of the four
corresponding computer algebra programs, is that each DE has to be
solvable for a leading derivative. Extra constraints for the
conservation laws can be specified. Examples include new conservation laws
that are non-polynomial in the functions, that have an explicit
variable dependence and families of conservation laws involving
arbitrary functions. The following equations are investigated in 
examples: Ito, Liouville, Burgers, Kadomtsev-Petviashvili,
Karney-Sen-Chu-Verheest, Boussinesq, Tzetzeica, Benney.
\end{abstract}

\section{Introduction}
As is well known, conservation laws play an important role in
mathematical physics. 
The knowledge of conservation laws is useful in the numerical integration
of partial differential equations (PDEs) 
\cite{LV}, for example, to control numerical errors.
Also, the investigation of conservation laws of the
Korteweg de Vries equation was the starting point of the discovery of
a number of techniques to solve evolutionary equations \cite{Ne}
(Miura transformation, Lax pair,
inverse scattering technique, bi-Hamiltonian structures). The existence
of a large number of conservation laws of a PDE (system) 
is a strong indication of its
integrability. Conservation laws play an important role in the theory of
non-classical transformations \cite{Mik1},\cite{Mik2} and in the
theory of normal forms and asymptotic integrability \cite{Mik4}.
Programs described below are able to find conservation laws involving
the independent variables explicitly. Finding such conservation laws
is a good challenge for the inverse scattering technique.

The purpose of the methods described below is to pose as few
restrictions as possible on the differential equations (DEs) to be
investigated. For example, 
it is not assumed that any Lie symmetries are known, nor that the equations
are equivalent to the Euler-Lagrange equations of a variational problem.
Instead we attempt to solve the conservation law condition directly.
The strategy will be to make a local ansatz involving only the 
dependent variables and their derivatives.
Further, the order of the derivatives is bounded
in order to obtain an over-determined PDE problem which subsequently is
solved with the computer algebra package {\sc Crack} \cite{CRACK1}, 
\cite{CRACK2}. 

In an earlier paper \cite{WBM} three of the methods were discussed
with emphasis put on the computer algebra algorithms involved. 
In this paper we present an additional fourth method and compare these
methods in terms of complexity and functionality.

The rest of the paper is organised as follows. After specifying the
notation that is used, in section \ref{sec2} a
reminder on issues of the  equivalence of conservation laws will
provide the motivation for the four approaches which are explained in
section \ref{sec3} followed by an overview. In section
\ref{sec9} related computer algebra programs are shortly described and
examples are given. Extensions of the basic usage of these programs
are discussed in section \ref{exapp}.

\section{Notation and setup}
We adopt the notation of the book of Olver \cite{PO2} where the
question of equivalence of conservation laws is described in more
detail in Chapter 4.3. This compact notation will be explained using
the sine-Gordon equation which will be used as an example throughout
the paper. 
\begin{itemize}
\item In general derivations or generally applicable formulas the
  independent variables are $x = (x^1, x^2, \ldots , x^p)$.
  In examples independent variables are $t,x$ and $y$.
\item Functions are denoted in general by $u = (u^1, u^2, \ldots, u^q)$.
  In examples the function is $u$ and in a few examples an additional
  function is $v$.
\item Partial derivatives are written as lower index as in
  $u_{tx} - \sin u = 0$.  If partial derivatives are repeated $n$ times
  ($n>2$) then this may be indicated by writing $n$ in front of the 
  variable in the index
  like in $\partial^7 u/(\partial t^4\partial x^3) = u_{4t3x}$ (which 
  is especially used in the appendix due to the high derivatives
  occuring there). 
\item An upper index in brackets like $u^{(n)}$ denotes the set of all
  derivatives of all components $u^1,\ldots,u^q$ of order up to $n$
  including order zero, i.e.\ the functions themselves. In our example
  with $u=u(t,x)$ this would, for example, mean 
  $u^{(2)} = \{u,u_t,u_x,u_{tt},u_{tx},u_{xx}\}$.
\item The differential equations that are to be investigated
  concerning conservation laws are denoted $0=\Delta(x,u^{(n)})$, in
  our example $\Delta(x,u^{(2)})=u_{tx}-\sin u$. If we have $q$
  functions $u^1,\ldots,u^q$ then it is assumed that a system of $q$
  equations $0=\Delta_1, 0=\Delta_2,\ldots,0=\Delta_q$ is given.
\item Latin indices $i,j$ have a range $1,\ldots,p$ used for $x^i$ and
  Greek indices $\alpha,\mu$ have a range $1,\ldots,q$ used for
  $u^\alpha$ and $\Delta_\mu$.
\item Whereas $u^{(n)}$ denotes the set of all possible indices up to
  order $n$ we will also need a way to specify any particular partial
  derivative of an unspecified order. For that we will use the
  so-called multiple index $_J$. Then any $u_x, u_t, u_{2t5x}$ are all
  examples for $u_J$. An example for the use of $_J$ is the notation
  of total derivatives:
\item For the total derivative we use the symbol $D$:
  \[D_{x^i} = \partial_{x^i} + \sum_{\alpha=1}^q \left(
  u^\alpha_{x^i}\partial_{u^\alpha} +
  \sum_{j=1}^p u^\alpha_{x^jx^i}\partial_{u^\alpha_{x^j}} + \ldots
  \right). \]
  Using the multi index $_J$ a more compact notation is:
  \[D_{x^i} = \partial_{x^i} + 
  \sum_{\alpha=1}^q \sum_J u^\alpha_{Jx^i} \partial_{u^\alpha_J}. \]
\item If a relation $0 = W(x,u^{(n)})$ is said to be satisfied
  identically for any solutions of the equation(s) $0=\Delta$ then
  this means that $W$ can be written as a linear combination of
  $\Delta$ and any total derivatives of $\Delta$ with arbitrary 
  $x-$ and $u^{(n)}-$dependent coefficients which are non-singular for
  $0=\Delta$. The notation adopted for that relation is 
  \[0 = \left. W \right|_{\Delta=0} \] which may be described as
  ``$W$ vanishes modulo $\Delta=0$''. The way to check this circumstance
  in practice is 
  \begin{itemize}
  \item to solve each  $\Delta_\mu$ for one leading derivative
    (such that no other derivative in $\Delta_\mu$ is a derivative of its
    leading derivative and that no two leading
    derivatives from two different $\Delta_\mu, \Delta_\nu$ coincide and
    that none is a derivative of another one),
  \item to substitute in $W(x,u^{(n)})$ all these eliminated leading
    derivatives as well as all derivatives of the leading derivatives.
  \item We have $0 = \left. W \right|_{\Delta=0}$ if and only if $W$ vanishes
    identically after all possible substitutions have been performed.
  \end{itemize}
  {\it Example:} For the sine-Gordon equation $0=\Delta=u_{tx} - \sin u$
  we will take $u_{tx}$ as leading derivative and do substitutions
  $u_{tx}=\sin u$. Let us assume an expression
  \begin{equation}
  W = -8u_{xx}u_{xxt}+4u_x^3u_{tx}+8u_xu_{xx}\cos u-4u_x^3\sin u \label{mod1}
  \end{equation}
  is given and we want to compute $W |_{\Delta=0}$. The highest
  derivative of $u_{tx}$ is $u_{xxt}=\partial_x u_{tx}$. We therefore
  substitute $u_{txx}=\partial_x(\sin u)= u_x\cos u$ giving
  \begin{equation}
  W = 4u_x^3u_{tx}-4u_x^3\sin u . \label{mod2}
  \end{equation}
  Substituting now $u_{tx} = \sin u$ gives $W=0$.
  We therefore found that $W$ as given in (\ref{mod1}) satisfies
  $W |_{\Delta=0}=0$.
\item A conservation law of $\Delta$ will be given in form of a
  so-called conserved current $P^i$ where each $P^i$ is a differential
  expression in $x$ and $u$, i.e.\ $P^i=P^i(x,u^{(m)})$ which has a
  vanishing divergence due to $0=\Delta$. In other words
  \[ 0 = \left. \sum_i D_{x^i} P^i \right|_{\Delta=0} 
       = \left. \mbox{Div}\,P \right|_{\Delta=0}.\]
  {\it Example:} For the current
  \begin{equation}
  P^t=-4u_{xx}^{\;\;2}+u_x^4, \;\;\;\;\;\; P^x=4u_x^2\cos u , \label{mod3}
  \end{equation}
  we find
  \begin{equation}
  \mbox{Div} P = D_t P^t + D_x P^x =
  -8u_{xx}u_{txx}+4u_x^3u_{tx}+8u_xu_{xx}\cos u -4u_x^3\sin u ,
  \label{mod4}
  \end{equation}
  which is equal $W$ in (\ref{mod1}). Because of
  $\mbox{Div}\,P|_{\Delta=0} = W|_{\Delta=0} = 0$ the vector
  $P$ in (\ref{mod3})
  is a conserved current of the sine-Gordon equation and represents
  a conservation law.
\item About the conserved quantity: By
  integrating $0=\mbox{Div}\,P$ over a region of the $p$-dimensional
  $x^i$-space we obtain the vanishing of a surface integral over that
  region: \[ 0 = \oint P^i dS_i .\]
  By choosing a region with a cylinder-like shape where the radius of
  the cylinder is very large and the bottom and
  top side of the cylinder lie in the $p-1$ dimensional surfaces
  $x^1=a=$ constant, and 
  $x^1=b=$ constant the surface integral takes the form
  \[0 = \int_{x^1=b} P^1 dx^2\ldots dx^p 
      - \int_{x^1=a} P^1 dx^2\ldots dx^p \]
  if we assume that $P^1$ falls off sufficiently quickly for any
  $x^i \rightarrow \infty$ and therefore the integral over the mantle
  of the cylinder vanishes
  $\int_{\mbox{\scriptsize mantle}} P^i dS_i \rightarrow 0$.
  The minus sign comes in because the normal vectors of the top and
  bottom surface point into opposite directions.

  If one of the coordinates plays the role of time, say $x^1$, then
  $\int P^1 dx^2\ldots dx^p$ is called a constant of motion as it does
  not change from any time $a$ to any time $b$.

  {\it Example:} Continuing our example we have as a constant of motion
  \[ \int_{-\infty}^\infty (-4u_{xx}^{\;\;2}+u_x^4) \, dx\]
  if $u$ is a solution of the sine-Gordon equation.
\end{itemize}

\section{The equivalence of conservation laws} \label{sec2}
Although two conservation laws may look rather different, i.e.\ their two
conserved currents, say $P$ and $\tilde{P}$, may be different, nevertheless
their information content may be the same.
To have a method of counting conservation laws and of
comparing them, we need a unique way to characterise them.
To do that we first look at ways conservation laws can look different
but be equivalent.
Two conservation laws \ \ $\mbox{Div}\,P = 0$ \ \ and \ \
$\mbox{Div}\,\tilde{P} = 0$ \ \ are equivalent
if $0 = \mbox{Div} (P-\tilde{P}) = \mbox{Div}\,R$ is a 
trivial conservation law.\vspace{6pt}

{\bf (i)} The first kind of equivalence of two conservation laws 
is the case that $R=0$ for all
solutions of $\Delta=0$, i.e.\ if $P$ and $\tilde{P}$ differ only by
multiples of $\Delta$ and by total derivatives of $\Delta$ (i.e.\ by
$D_J\Delta$).  To test whether this is the case for two given
conserved currents $P$ and $\tilde{P}$ one has to check whether 
\[0 = \left.(P - \tilde{P}) \right|_{\Delta=0} \] holds.
As described above one solves the equation $0=\Delta$
(or system of equations $0=\Delta_\mu$)
for the leading derivative(s) and substitutes that (them) 
in $P - \tilde{P}$. If $P - \tilde{P}$ becomes identically zero then
the two conservation laws based on $P$ and $\tilde{P}$ are equivalent.

If the conservation laws are not yet calculated and one wants to 
ensure that the computation of $P$ gives a unique result, without
arbitrariness due to terms vanishing because of $\Delta=0$ then
there is no need to solve $\Delta = 0$ for some leading 
derivative(s) $u_J$. In that case one just drops from the
beginning of the calculation the dependency of $P$ on
the leading derivative(s) $u_J$ and all derivatives of $u_J$.

{\em Example:} When computing conserved currents $P$ for the sine-Gordon
equation $0 = u_{tx} - \sin u$, then in the ansatz for $P$ the
components $P^t, P^x$ are assumed to be independent of $u_{tx}$
and derivatives of $u_{tx}$, like $u_{ttx}, u_{txx}, \ldots$
(see the appendix).\vspace{6pt}

{\bf (ii)} The second kind of equivalence of two conservation
laws occurs if  
$R^i = P^i - \tilde{P}^i = \sum_j D_j V^{ij}$ for some
expressions $V^{ij}(x,u^{(m)})=-V^{ji}$, anti-symmetric in $i,j$,
because in that case
$\mbox{Div}\,R = \sum_{i,j} D_i D_j V^{ij} = 0$
(due to the symmetry $D_iD_j=D_jD_i$ and the anti-symmetry $V^{ij}=-V^{ji}$)
for any functions $u$, not necessarily solutions $u(x)$ of $\Delta=0$.
The existence of $V^{ij}(x,u^{(m)})$ satisfying
$P^i = \tilde{P}^i + \sum_j D_j V^{ij}$ may not be obvious and may
require a computation checking $\mbox{Div}\,P = \mbox{Div}\,\tilde{P}$.
For ODEs this problems does not occur as
there is only one independent variable and no antisymmetric $V^{ij}$.

The solution to this problem is not to compare conservation laws by
comparing their conserved currents $P$ and $\tilde{P}$ 
but by comparing them by their
integrating factors, for PDEs they are called characteristic
functions, in the following way. For a conservation law to satisfy
$\left. \mbox{Div}\,P \right|_{\Delta=0} = 0$
means that $\mbox{Div}\,P$ is identical to a linear combination of
$\Delta_\mu$ and total derivatives $D_J\Delta_\mu$. Partial integration 
can rewrite that as a
divergence plus a linear combination of the $\Delta_\mu$ alone:
\begin{eqnarray}
\left. \mbox{Div}\,P \right|_{\Delta=0} & = & 0  \label{trafo0} 
\\ \Longleftrightarrow \;\;
\exists Q^J_\nu: \;\;\mbox{Div}\,P & = & \sum_{\nu,J} Q^J_{\nu}
D_J\Delta_{\nu} \;\;\;\;\;\;\;\;\;\;\;\;\;\;\;\;\;\;\;\;\;\;\;\;\;\,
(\mbox{identically in {\it all}} \;\; x, u^{\alpha}_J ) \label{trafo1} \\
& = & \sum_{\nu,J} D_J(Q^J_{\nu} \Delta_{\nu}) - D_J(Q^J_{\nu}) \Delta_{\nu}
\;\;(\mbox{repeated partial integration}) 
\nonumber \\
& = & \mbox{Div}\,R + \sum_{\nu} Q_{\nu} \Delta_{\nu}  \nonumber \\
\Longleftrightarrow \;\;\;
\mbox{Div}\,P & = & \sum_{\nu} Q_{\nu} \Delta_{\nu} \;\;\;\;
\;\;\;\;\;\;\;\;\;\;\;\;\;\;\;\;\;\;\;\;\;\;\;\;\;\;
(\mbox{after renaming $(P\!-\!R) \rightarrow P$}). \label{trafo3}
\end{eqnarray}
The integrating factors $Q_{\nu}$ are called characteristic functions
as it is known (\cite{PO2}, p.\ 272) that for a totally
non degenerate system $\Delta=0$, the equivalence class of conservation laws
$\mbox{Div}\,P|_{\Delta=0} = 0$ is characterised uniquely by the functions
$Q_{\nu}$ up to equivalence of type {\bf (i)}.

One can look at equation (\ref{trafo3}) as a determining equation for
$P$ and $Q_\nu$ as in the method described in section
\ref{sec5} below. Alternatively one can formulate a system of
conditions 
that is equivalent to (\ref{trafo3}) but which involves only functions
$Q_\nu$. That is achieved using the property of
Euler operators (also called variational derivatives)
$E_{\nu} = \sum_J (-D)_J \partial/\partial u^{\nu}_J$ when acting
on an expression they give identically zero iff this expression 
is a divergence. 
Conditions for the $Q_{\nu}$ are therefore
\begin{equation}
 \forall \nu: \;\;\;\;
  0 = E_{\nu} \left( \sum_{\mu} Q_{\mu} \Delta_{\mu} \right)
    = \sum_J (-D)_J
        \left( \frac{\partial}
                    {\partial u^{\nu}_J} \sum_{\mu}Q_{\mu} \Delta_{\mu}
        \right).                                        \label{c4}        
\end{equation}
{\em Example:} Allowing $Q$ to be of $2^{\mbox{\scriptsize nd}}$ order,
i.e.\ $Q=Q(t,x,u,u_t,u_x,u_{tt},u_{xx})$ where we dropped the
dependence on $u_{tx}$ which is equal to $\sin u$ (see remark above) we
find that \[Q\cdot\Delta=Q\cdot(u_{tx}-\sin u)\] does depend on 
$u,u_t,u_x,u_{tt},u_{tx},u_{xx}$ and the Euler operator therefore reads
\begin{eqnarray*}
 E & = & \partial_u - D_t\partial_{u_t} - D_x\partial_{u_x} +
(-D_t)(-D_t)\partial_{u_{tt}} + (-D_t)(-D_x)\partial_{u_{tx}} + 
(-D_x)(-D_x)\partial_{u_{xx}} \\
 & = & \partial_u - D_t\partial_{u_t} - D_x\partial_{u_x} +
D_t^2\partial_{u_{tt}} + D_tD_x\partial_{u_{tx}} +
D_x^2\partial_{u_{xx}}. 
\end{eqnarray*}

Requiring condition(s) (\ref{c4}) to be satisfied identically in all $x^i,
u^\alpha$ and derivatives of $u^\alpha$ (i.e.\ $u^\alpha_J$) is
equivalent to the condition (\ref{trafo3}). System (\ref{c4}) is often
very large. It can be considerably shortened if it is projected onto
the space of solutions $|_{\Delta=0}$ (as described above):
\begin{equation}
0 = \left. \sum_{\mu,J} (-D)_J
   \left( Q_{\mu} \frac{\partial \Delta_{\mu}}
                       {\partial u^{\nu}_J}
   \right)\right|_{\Delta=0} \;\;\; \forall \nu.  \label{c2}
\end{equation}
Conditions (\ref{c2}) are known as adjoint symmetry conditions which are
necessary but not sufficient for the $Q_{\mu}$ to be characteristic functions
of first integrals.\vspace{6pt}

{\bf iii)} For any two conservation laws 
$0=\mbox{Div}\,P$ and $0=\mbox{Div}\,\tilde{P}$,
$0=\mbox{Div}(P+\tilde{P})$ is also a conservation law. 
By determining conservation laws with
characteristic functions of successively increasing order, constant multiples
of characteristic functions of lower order can be dropped.\vspace{6pt}

{\bf iv)} In the case of (systems of) ODEs the characteristic functions are called
{\it integrating factors}, and $P$ is a scalar, called a {\it first
integral}. Any arbitrary function of first integrals is a first integral
as well. \vspace{6pt}

The four approaches described in the following four sections 
are to solve conditions (\ref{trafo0}), (\ref{trafo3}),
(\ref{c4}) and (\ref{c2}).
\section{The four approaches} \label{sec3}
\subsection{A first approach} \label{sec4}
The first approach is to solve 
\begin{equation} 
\left. \mbox{Div}\,P \right|_{\Delta=0} = 0 \label{a1}
\end{equation} 
directly. 

The condition (\ref{a1}) is made over-determined
by restricting the $P^i$ to be differential expressions 
in the $u$ of at most some order $k$, i.e.\ $P^i = P^i(x,u^{(k)})$. 
Characteristic features of this approach are
\begin{itemize}
\item[$(+)$] A single, first order PDE involving only few terms is to
be solved. 
\item[$(\mbox{0})$] Characteristic functions have to be computed from $P$
in a straightforward calculation (described in \cite{WBM}). This is
done within the computer algebra program
{\sc ConLaw1}  which implements the first approach
including the computation of related characteristic functions $Q_\mu$.
\item[$(-)$] It would be computationally expensive for a corresponding
computer program to drop {\em during the process of solving} (\ref{a1}) any
free functions $V^{ij}=-V^{ji}$ (see the discussion of the second kind
of equivalence of conservation laws in the previous section)
which correspond to trivial conservation laws\footnote{An algorithm
for that is given in \cite{WBM}.}. 
Hence, the condition (\ref{a1}) has to be solved first in full
generality and trivial conservation laws (i.e.\ $V^{ij}$)
have to be identified and dropped afterwards. That means that the task
for the computer program is made unnecessarily hard by the presence of
the trivial conservation laws in the general solution of (\ref{a1}). A
rule of thumb says that the difficulty in solving 
a linear over determined PDE system depends less on the order or
size of the PDE but more on the complexity of the result\footnote{For
example, if an over-determined PDE (system) has no solution then a
differential Gr\"{o}bner Basis calculation will quickly produce PDEs of
lower and lower order until a contradiction is reached. On the other
hand, if a PDE system has arbitrary functions in its general solution
(as is the case with the PDE (\ref{a1})) then computing a differential
Gr\"{o}bner Basis will {\it not} produce a system that is solvable by
only integrating ODEs, it will involve PDEs.}. 
That means the trivial conservation laws will complicate the solution
of (\ref{a1}), the more so the more independent variables are present. 
\item[$(-)$] In most cases the expressions for the $P^i$ are more
complicated than the expressions for the characteristic functions $Q_{\mu}$
which by the above rule of thumb indicates a more difficult computation than
the solution of equations involving only $Q_{\mu}$.
\end{itemize}
To illustrate and compare all four approaches we will apply each to finding
conservation laws of the sine-Gordon equation
\begin{equation} u_{tx} - \sin(u) = 0. \label{sG} \end{equation}
If the program {\sc ConLaw1}
is called to find conservation laws with conserved
current $P^t, P^x$ of order 0, then it will reply that it is not
applicable. This is because Div $P$ would be of first order in $u$, so
equation (\ref{sG}) could not be used to substitute $u_{tx}$ 
(when computing $|_{\Delta=0}$ in (\ref{a1})) and
therefore any conservation laws found would be valid for any function
$u(t,x)$, not necessarily only for solutions of the sine-Gordon
equation. These conservation laws would therefore be trivial,
falling into category {\bf (ii)} in section \ref{sec2}.

Details of higher order investigations 
are given in table 1 below. $u^{(n)}$ stands for all
derivatives of $u$ of order 0 to $n$. $u_{tx}^{\;\,(n)}$ stands for all
derivatives of $u_{tx}$ up to order $n$, for example, $u_{tx}^{\;\,(1)}$
would be the derivatives $u_{tx}, u_{ttx}, u_{txx}$.
Finally, $u^{(n)}/u_{tx}^{\;\,(k)}$ stands for all derivatives of $u$ up to order
$n$ apart from $u_{tx}$ and all its derivatives up to order $k$.
The conservation laws are given in the appendix (in the table only the
equation number is cited). For each conservation law in the
appendix (apart from the first) there exists another one resulting from
the exchange $t \leftrightarrow x$.

The times given in the table are measured on a 266 MHz Pentium PC
running a 80 MByte {\sc Reduce} 3.6 session under {\sc Linux} using the
Sep.\ 1998 version of the program {\sc Crack} for solving
the over-determined conditions. The 80 MByte
were not necessary. For example, it is possible (using {\sc ConLaw2} 
which implements the $4^{\mbox{\scriptsize th}}$ method described below) to
investigate up to $4^{\mbox{\scriptsize th}}$ order laws with 4 MByte and
up to $7^{\mbox{\scriptsize th}}$ order laws with 8 MByte.
To get this high in order with relatively low memory consumption,
one has to give  in {\sc Crack}
the study of integrability conditions a higher priority than
the integration of equations. 
The price is a higher computing time.
The times in the last column are to be understood only as {\it very} rough
indicators\footnote{For example, the computing times reported in
\cite{WBM} are at the time of revision of this paper (July 1999) 
already reduced by a factor of more than ten for higher orders.}.
They depend sensitively on the order of priorities with which modules 
are to be used within the program {\sc Crack} (see the manual
\cite{CRACK1} and about its availability
the end of the section \ref{summary}). 

When condition (\ref{a1}) is solved, the $P^i$ that are
computed initially do not contain $u_{tx}$ nor its derivatives.
Afterwards a computation as outlined in (\ref{trafo0}) -
(\ref{trafo3}) is performed such that finally {\sc ConLaw1}
is able to return the conservation law in the form (\ref{trafo3}).
In the process of computing this form (\ref{trafo3})
the new $P^i$ may now involve $u_{tx}$ (through 
$R$ in (\ref{trafo1}) - (\ref{trafo3})). 

\begin{center}
\begin{tabular}{|c|c|c|c|c|c|} \hline
 order & no of & independent       & functions to              & cons.       & time to      \\ 
of $P^i$ & terms & variables, [no of var.]& compute, [no of arg.]& laws        & solve (\ref{a1}) \\ 
 & & & & found & \\ \hline
 & & & & & \\
 1  &   8    & $t,x,u^{(2)}/u_{tx}$, [7]  & $P^t,P^x(t,x,u^{(1)})$,       [5]   & (\ref{l1}),(\ref{l2}) & 9 sec\\
 & & & & & \\ \hline
 & & & & & \\
 2  &  12    & $t,x,u^{(3)}/u_{tx}^{\;\,(1)}$, [9]& $P^t,P^x(t,x,u^{(2)}/u_{tx})$, [7] & (\ref{l3}) & 38 sec\\
 & & & & & \\ \hline
 & & & & & \\
 3  &  18    & $t,x,u^{(4)}/u_{tx}^{\;\,(2)}$, [11]& $P^t,P^x(t,x,u^{(3)}/u_{tx}^{\;\,(1)})$, [9]  & 
                 none\footnotemark\ & - \\
 & & & & & \\ \hline
 & & & & & \\
 4  &  26    & $t,x,u^{(5)}/u_{tx}^{\;\,(3)}$, [13]&
 $P^t,P^x(t,x,u^{(4)}/u_{tx}^{\;\,(2)})$, [11]& 
                 none\footnotemark\ & - \\
 & & & & & \\ \hline
\end{tabular} 
\begin{tabbing}
Table 1: \= The program {\sc ConLaw1} applied to compute conservation 
            laws of the \\
         \> sine-Gordon equation.
\end{tabbing}
\end{center}
\addtocounter{footnote}{-1}
\footnotetext{{\sc Crack} was not able to solve all the equations completely
              because the general solution of (\ref{a1}) involves free
              functions (related to
              trivial conservation laws) which complicates the problem
              considerably for the computer program.}
\addtocounter{footnote}{1}
\footnotetext{The computer memory was not sufficient to complete the computation.}
\subsection{A second approach}\label{sec5}
The next approach consists in solving
\begin{equation}
\mbox{Div}\,P = \sum_{\nu} Q_{\nu} \Delta_{\nu}  \label{a3}
\end{equation}
directly, i.e.\ finding $P^i, Q_{\mu}$ that satisfy (\ref{a3})
identically in $x^i, u^{\alpha}_J$. Equations $\Delta=D_J\Delta=0$ are
{\it not} used for substitutions in (\ref{a3}) but they are used to reduce
dependencies of the $Q_{\mu}$.

The problem becomes over-determined by restricting the order of the
$Q_{\mu}$, i.e.\ $Q_{\mu}=Q_{\mu}(x,u^{(k)})$ for some $k$ and by
taking $Q_\mu \leftarrow Q_\mu |_{\Delta=0}$, i.e.\
having $Q_\mu$ independent of one leading $u$-derivative (and their derivatives)
from each one of the equations $\Delta_{\nu}$.
If $Q_{\mu}$ would be allowed to depend
on all $u^{(n)}$ which occur in (\ref{a3})
then this equation could
simply be solved algebraically, by eliminating one of the $Q_{\mu}$. But
that would mean division through one $\Delta_{\mu}$ and therefore $Q_{\mu}$ being
singular for solutions of $\Delta_{\mu}=0$. 



The second approach has the following characteristics:
\begin{itemize}
\item[$(+)$] The conservation law condition (\ref{a3}) 
             is a single first order PDE as in the
             first approach.
\item[$(+)$] By calculating characteristic functions $Q_\mu$ and 
             furthermore characteristic functions $Q_\mu |_{\Delta=0},$
             conservation laws are uniquely characterized.
\item[$(+)$] The effort in formulating conditions is as low as in the first
             approach.
\item[$(\mbox{0})$] The $P^i$ and $Q_{\mu}$ are computed in one computation.
\item[$(\mbox{0})$] The number of functions to compute is higher than
             in the first approach and also the number of derivatives of
             $u$ on which these functions depend on because no substitutions
             are done in (\ref{a3}). The resulting complication is not
             too big as more variables means a higher over
             determination and simplification. 
\item[$(-)$] If the order of $\Delta$ is $n$ and the order of $Q_{\mu}$ is
             chosen to be $k$ then the order of $P^i$ at the start of the
             computation can be assumed without loss of generality to
             be $\mbox{max}(k,n)$ (see \cite{PO1}). `Without loss of
             generality' means that a trivially conserved current
             $\tilde{P}$ can be subtracted from $P$ such that
             $P-\tilde{P}$ is of order $\mbox{max}(k,n)$. If the right-hand
             side of equation (\ref{a3}) is known to be linear in the
             highest derivatives of order $\mbox{max}(k,n)$ then
             $P^i$ at the start of the
             computation can even be assumed without loss of generality to
             be of order $\mbox{max}(k,n)-1$.

             In this approach the investigations with
             $k<n$ are not much simpler than the case $k=n.$ This matters
             when the order $n$ of $\Delta$ and the number $p$ of variables
             $x$ are high. 
             Therefore this approach is not very efficient for low
             order conservation laws of high order equations.

             For example, for zeroth order conservation laws $(k=0)$
             of the Kadomtsev-Petviashvili equation
             (\ref{KP}) the $P^i$ are taken
             initially as functions of the 23 variables $t,x,y,$
             $u,u_t,$ $u_x,u_y,$
             $u_{tt},\ldots,u_{yy},$ $u_{ttt},\ldots,u_{yyy}$ and the 
             conservation law condition
             (\ref{a3}) is a condition in 38 variables (including the $4^{\mbox{\scriptsize th}}$
             order $u$-derivatives).
             That is a much harder problem than the corresponding
             conditions (\ref{c4}),(\ref{c2}). For example, in this
             case condition (\ref{c4}) is a single $4^{\mbox{\scriptsize th}}$ order PDE in 
             also 38 variables
             but for only {\it one} function $Q$ of only {\it four} variables!
\item[$(-)$] When looking for conservation laws with the first method, gradually
             increasing the order of the conserved current $P$ gives each
             conservation law in its lowest order form, i.e.\ a form where
             $P$ is of minimal order. This is not necessarily the case using
             the $2^{\mbox{\scriptsize nd}}$ method. The transformation 
             (\ref{trafo1})-(\ref{trafo3}) adding $R$ to
             $P$ may increase the order of $P$. This implies an
             increase of complexity having to go up in order to get
             the equivalent conservation law. To give an example, 
             the Tzetzeica equation
             $u_{xt}= e^u - e^{-2u}$ (analysed in
             \cite{SZ},\cite{Mik3}) has the conservation law
\begin{eqnarray*}
0 & = & D_t\left[3u_{xxx}^{\;\,2} - 5u_{xx}^{\;3} +
                 15u_{xx}^{\;3}u_x^2 + u_x^6\right ]   +  \\
  &   & D_x\left[- 3e^u\left( u_{xx}^{\;2} + 2u_{xx}u_x^2 + 2u_x^4 \right)
                 - 3e^{-2u}\left( 2u_{xx}^{\;2} - 8u_{xx}u_x^2 +
                 u_x^4 \right)\right]
\end{eqnarray*}
with a third order conserved current. (In \cite{Mik3} an infinite list
of conservation laws is given.)
Bringing the above conservation law to the form (\ref{a3})
as it would be found with the second method, it becomes
\begin{eqnarray*}
& \;\; & 6\left( u_{xxxxx} + 5u_{xxx}u_{xx} - 5u_{xxx}u_x^2 -
       5u_{xx}^{\;2}u_x + u_x^5 \right)
      \left( u_{tx} - e^u + e^{-2u} \right)   \\
& = & D_t\left[3u_{xxx}^{\;\,2} - 5u_{xx}^{\;3} +
               15u_{xx}^{\;2}u_x^2 + u_x^6\right] +  \\
& \;\;  & D_x \,3\left[2u_{tx}u_{xxxx} - 2u_{txx}u_{xxx} + 5u_{tx}u_{xx}^{\;2}
       - 10u_{tx}u_{xx}u_x^2          \right.      \\
& \;\;  & \;\;\;\;\;\;\;\;\; + e^u\left(2u_{xxx}u_x - 2u_{xxxx} - 6u_{xx}^{\;2}
                  + 8u_{xx}u_x^2 - 2u_x^4\right)   \\
& \;\;  & \;\;\;\;\;\;\;\;\, \left. + e^{-2u}\left(2u_{xxxx} +
          4u_{xxx}u_x + 3u_{xx}^{\;2} - 2u_{xx}u_x^2 - u_x^4\right)
         \right] 
\end{eqnarray*}
with a $4^{\mbox{\scriptsize th}}$ order conserved current.
\end{itemize}
Applying the program {\sc ConLaw3} that corresponds
to the above method to the sine-Gordon equation (\ref{sG})
gives the following table.
\begin{center}
\begin{tabular}{|c|c|c|c|c|c| } \hline
 order & no of & independent       & functions to                & cons. & time to      \\ 
of $Q$ & terms & variables, [no of var.] & compute, [no of arg.] & laws  & solve (\ref{a3}) \\
 & & & & found & \\ \hline
 & & & & & \\
 0  &  10    & $t,x,u^{(2)}$, [8]&   $P^t,P^x(t,x,u^{(1)})$, [5]    & none        & 3 sec\\ 
    &        &          &    $Q(t,x,u)$, [3]      &             &  \\ \hline
 & & & & & \\
 1  &  10    & $t,x,u^{(2)}$, [8]& $P^t,P^x,Q(t,x,u^{(1)})$, [5]     & (\ref{l1}),(\ref{l2}) & 8.3 sec\\ 
    &        &          &                        &  &       \\ \hline
 & & & & & \\
 2  &  10    & $t,x,u^{(2)}$, [8]& $P^t,P^x(t,x,u^{(2)})$,   [8]   & none         & 3.5 sec\\ 
    &        &        & $Q(t,x,u^{(2)}/u_{tx})$, [7] &              &            \\ \hline
 & & & & & \\
 3  &  16    & $t,x,u^{(3)}$, [12]& $P^t,P^x(t,x,u^{(3)})$,   [12]  & none\footnotemark\  & - \\
    &        &        & $Q(t,x,u^{(3)}/u_{tx}^{\;\,(1)})$, [9] &  & \\ \hline
\end{tabular}
\begin{tabbing}
Table 2: \= The program {\sc ConLaw3} applied to compute conservation 
            laws of the \\
         \> sine-Gordon equation.
\end{tabbing}
\end{center}
\footnotetext{The computer memory was not sufficient to complete the computation.}
\subsection{A third approach}
Instead of calculating the conserved current $P^i$ directly, the third
approach is to calculate characteristic functions $Q_{\mu}$ first
(see, e.g. Proposition 5.33 in \cite{PO2})
and from them $P^i$ afterwards using formulas of Anco \& Bluman
\cite{AB1},\cite{AB2},\cite{AB4} 
in a form described in \cite{WBM} or using repeatedly the {\sc Crack}
routine for integrating exact DEs (see also the section on Homotopy
Operators in \cite{PO2}). The condition (\ref{c4})
(as derived in \cite{PO2},\cite{AB1}) is: 
\begin{equation}
  0 = \sum_J (-D)_J
        \left( \frac{\partial}
                    {\partial u^{\nu}_J} \sum_{\mu}Q_{\mu} \Delta_{\mu}
        \right)\;\;\;\; \forall \nu.                 \label{c4a}       
\end{equation}
Typical features are:
\begin{itemize}
\item[$(+)$] Equations (\ref{c4a}) are equivalent to (\ref{a3}) and
             therefore necessary and sufficient.
\item[$(+)$] The usually more complicated $P^i$ are eliminated and as
             in the $2^{\mbox{\scriptsize nd}}$ method, no trivial conservation laws are
             calculated which otherwise unnecessarily complicate the
             calculation.
\item[$(+)$] The highest $u$-derivatives in conditions
          (\ref{c4a}) are of the order $2n$ where $n$ is the order of the
          $u$-derivatives in $\sum_{\mu}Q_{\mu} \Delta_{\mu}$.
          The harder the problem, i.e.\ the higher $n$ and the
          higher the number of 
          variables, the more $u$-derivatives occur only explicitly in
          (\ref{c4a}) and can be used for a direct separation
          (splitting). Higher
          over determination simplifies the solution of (\ref{c4a}).
\item[$(-)$] Equations (\ref{c4a}) consist of as many equations as there are
             dependent variables $u^{\mu}$. The unknown functions
             $Q_{\mu}$ appear with $n^{\mbox{\scriptsize th}}$ 
             order derivatives.
\item[$(-)$] For an increasing order of the $Q_{\mu}$,  increasing number of
             $u^{\nu}$ and increasing number of $x^i$, the size of
             (\ref{c4a}) can soon become unmanageable. 
\end{itemize}
Applying the program {\sc ConLaw4} that corresponds
to the above method to the sine-Gordon equation (\ref{sG})
gives the following table 3. The striking feature of this approach is
the quick increase of the ``size of conditions''. 
Apart from the order 0 case they increase by a factor of about 7 which
itself is increasing slightly with the order. The size of conditions
prevents going higher in the order. On the other hand, the
completeness of the generated conditions simplifies the solution in
difficult cases and speeds up the solution of the over determined
system as long as it is not already too large at the beginning.
\begin{center}
\begin{tabular}{|c|c|c|c|c|c| } \hline
 order & no of & independent       & functions to                & cons.  & time to      \\ 
       & terms & variables, [no of var.] & compute, [no of arg.] & laws   & solve (\ref{c4a}) \\
 & & & & found & h:min:sec \\ \hline
 & & & & & \\
 0  &   7    & $t,x,u^{(1)},u_{tx}$, [6]& $Q(t,x,u)$, [3]               & none       & 0.7 sec\\ 
 & & & & & \\ \hline
 & & & & & \\
 1  &  22    & $t,x,u^{(2)}$, [8]& $Q(t,x,u^{(1)})$, [5]                & (\ref{l1}) & 2.8 sec\\
 & & & & & \\ \hline
 & & & & & \\
 2  &  154   & $t,x,u^{(3)}$, [17]& $Q(t,x,u^{(2)}/u_{tx})$, [7]        & none       & 4.7 sec\\
 & & & & & \\ \hline
 & & & & & \\
 3  &  1116  & $t,x,u^{(4)}$, [24]& $Q(t,x,u^{(3)}/u_{tx}^{\;\,(1)})$, [9]  & (\ref{l3}) & 5 min 17 sec \\ 
 & & & & & \\ \hline
 & & & & & \\
 4  &  8402  & $t,x,u^{(5)}$, [34]& $Q(t,x,u^{(4)}/u_{tx}^{\;\,(2)})$, [11] & none & 10 h 49 min\footnotemark\ \\
 & & & & & \\ \hline
 & & & & & \\
 5  &  64064  & $t,x,u^{(6)}$, [41]& $Q(t,x,u^{(5)}/u_{tx}^{\;\,(3)})$, [13] & - & $>2$ days \\
 & & & & & \\ \hline
\end{tabular}
\begin{tabbing}
Table 3: \= The program {\sc ConLaw4} applied to compute conservation 
            laws of the \\
         \> sine-Gordon equation.
\end{tabbing}
\end{center}
\footnotetext{This time was nearly completely spent to formulate
the condition and to separate it into 823 individual equations for
$Q$. Then already the $3^{\mbox{\scriptsize rd}}$ step gave that $Q$ cannot
depend on $4^{\mbox{\scriptsize th}}$ order derivatives.}
\subsection{A fourth approach} \label{sec7}
Projecting conditions (\ref{c4a}) into the space of solutions of $\Delta=0$ we obtain
\begin{equation}
0 = \left. \sum_{\mu,J} (-D)_J
   \left( Q_{\mu} \frac{\partial \Delta_{\mu}}
   {\partial u^{\nu}_J}
   \right)\right|_{\Delta=0} \;\;\;\;\; \forall \nu.  \label{a2}
\end{equation}
The characteristic features of this method are similar to 
those of the third method with the following modifications:
\begin{itemize}
\item[$(+)$] The conditions usually involve fewer terms than in the
          third approach which can be decisive. But as the conditions
          (\ref{a2}) are not sufficient, they are less over determined
          and may be harder to solve than those in the third approach.
\item[$(-)$] Because the conditions (\ref{a2}) investigated by this fourth
          method are not equivalent to (\ref{c4a}) and sometimes less
          restrictive than conditions (\ref{c4a}), the solutions of
          (\ref{a2}) need not represent conservation laws. Therefore
          after computing the $Q_{\mu}$ from (\ref{a2}), it has to be
          checked whether $P^i$ exist that satisfy
          $\mbox{Div}\,P = \sum_{\nu} Q_{\nu} \Delta_{\nu}$
          (\cite{AB1},\cite{AB2},\cite{AB4},\cite{WBM}). 
          If they do not exist then the $Q_{\mu}$ correspond 
          to an adjoined symmetry but not to a conservation law.
\item[$(-)$] If the fourth method finds adjoined symmetries which are
          not conservation laws, then it is still possible that these
          adjoined symmetries can be combined linearly to give conservation
          laws. But how to combine adjoined symmetries to give
          conservation laws is not answered by solving (\ref{a2}), it
          has to be investigated separately. This will be illustrated
          with the following simple example.

          Applying the third method through the program {\sc ConLaw4}
          to the ODE $u''+u=0$ and restricting the search to
          integrating factors $Q=a(x)u + b(x)$ that are linear in $u$,
          the program finds the following 5 integrating factors 
          \[ \cos(x)^2 u' + \cos(x)\sin(x)u, \;\;
          2\cos(x)^2u - 2\cos(x)u'\sin(x) - u, \;\;
          \cos(x), \;\;
          \sin(x), \;\;
          u', \]
          and therefore it finds 5 first integrals. In comparison,
          using the same ansatz for $Q$ but now using the fourth
          method with the less restrictive condition (\ref{a2}) the
          corresponding program {\sc ConLaw2} finds 8 solutions for $Q$:
          \[ - 2\cos(x)^2u' - 2\cos(x)\sin(x)u + u', \;\;
          \cos(x), \;\;
          \sin(x), \;\;
          u', \]
          \[\cos(x)u'u + \sin(x)u^2, \;\;
          \cos(x)u^2  - u'\sin(x)u, \;\;
          - \cos(x)^2u + \cos(x)u'\sin(x), \;\;
          u. \]
          Only for the first 4 of these 8 $Q$-values does a $P$ exist
          such that $D_xP = Q\cdot(u''+u)$, i.e.\ only 4 first integrals
          are found, the remaining 4 solutions represent only adjoined
          symmetries. This inability to find 5 first integrals is not
          a weakness of the computer program but the price to pay for
          the fourth method to have shorter conditions (\ref{a2})
          compared with the conditions (\ref{c4a}) of the third method.
          If the conditions (\ref{a2}) of the fourth method are less
          restrictive than the conditions (\ref{c4a}) of the third
          method, why then does it find {\em fewer} first integrals,
          4 instead of 5? The answer is that the $5^{\mbox{\scriptsize th}}$
          conservation law is contained in the second half of the 8
          solutions: the $8^{\mbox{\scriptsize th}}$ solution 
          for $Q$ plus 2 times the $7^{\mbox{\scriptsize th}}$ solution gives a
          value for $Q$ that is not only an adjoined symmetry but also
          an integrating factor for an additional first integral.

          To summarise, the fourth method gives shorter, more
          manageable conditions which, strictly speaking, do not
          compute conservation laws but adjoined symmetries.
          Conservation laws can be derived through appropriate linear
          combinations of adjoined symmetries which has to be
          investigated separately. This theoretical weakness of the
          fourth method does usually play no role in practical
          applications. 
\end{itemize}
Applying the program {\sc ConLaw2} that corresponds
to the above method to the sine-Gordon equation (\ref{sG})
gives the following table. The typical feature of this approach is
the slower increase of the size of conditions. 
Apart from the order 0 case they increase by a factor of about 2 which
itself is increasing slightly with the order. Compared with the
previous method the size of conditions grows slower which allows to go
higher in the order. Because the conditions that are generated are only
necessary, not sufficient, they are slightly more difficult and
expensive to solve. This causes longer running times for low order
investigations. Time limitations could be overcome to some extend by
faster computers.
\begin{center}
\begin{tabular}{|c|c|c|c|c|c| } \hline
 order & no of & independent           & functions to          & cons. & time to      \\ 
 of $Q$ & terms & variables, [no of var.]& compute, [no of arg.]& laws  & solve (\ref{a2}) \\ 
 & & & & found & h:min:sec \\ \hline
 & & & & & \\
 0  &   6    & $t,x,u^{(1)}$, [5]&   $Q(t,x,u)$, [3]    & none        & 1 sec\\ 
 & & & & & \\ \hline
 & & & & & \\
 1  &  21    & $t,x,u^{(2)}/u_{tx}$, [7]& $Q(t,x,u^{(1)})$, [5]      & (\ref{l1}) & 4.3 sec\\ 
 & & & & & \\ \hline
 & & & & & \\
 2  &  45    & $t,x,u^{(3)}/u_{tx}^{\;\,(1)}$, [9]& $Q(t,x,u^{(2)}/u_{tx})$, [7]   & none  & 12 sec \\ 
 & & & & & \\ \hline
 & & & & & \\
 3  &  99    & $t,x,u^{(4)}/u_{tx}^{\;\,(2)}$, [11]& $Q(t,x,u^{(3)}/u_{tx}^{\;\,(1)})$, [9] & (\ref{l3}) & 50 sec\\
 & & & & & \\ \hline
 & & & & & \\
 4  &  202   & $t,x,u^{(5)}/u_{tx}^{\;\,(3)}$, [13]& $Q(t,x,u^{(4)}/u_{tx}^{\;\,(2)})$, [11] & none & 2 min 43 sec\\
 & & & & & \\ \hline
 & & & & & \\
 5  &  435   & $t,x,u^{(6)}/u_{tx}^{\;\,(4)}$, [15]& $Q(t,x,u^{(5)}/u_{tx}^{\;\,(3)})$, [13] & (\ref{l4}) & 16 min 10 sec\\
 & & & & & \\ \hline
 & & & & & \\
 6  &  870   & $t,x,u^{(7)}/u_{tx}^{\;\,(5)}$, [17]& $Q(t,x,u^{(6)}/u_{tx}^{\;\,(4)})$, [15] & none  & 49 min 20 sec\\
 & & & & & \\ \hline
 & & & & & \\
 7  &  1836  & $t,x,u^{(8)}/u_{tx}^{\;\,(6)}$, [19]& $Q(t,x,u^{(7)}/u_{tx}^{\;\,(5)})$, [17] &  (\ref{l5}) & 8 h\\
 & & & & & \\ \hline
 & & & & & \\
 8  &  3643  & $t,x,u^{(9)}/u_{tx}^{\;\,(7)}$, [21]& $Q(t,x,u^{(8)}/u_{tx}^{\;\,(6)})$, [19] & none  & 5 h 22 min\\
 & & & & & \\ \hline
 & & & & & \\
 9  &  7434  & $t,x,u^{(10)}/u_{tx}^{\;\,(8)}$, [23]& $Q(t,x,u^{(9)}/u_{tx}^{\;\,(7)})$, [21] &  (\ref{l6}) & 25 h\\
 & & & & & \\ \hline
 \end{tabular}
\begin{tabbing}
Table 4: \= The program {\sc ConLaw2} applied to compute conservation 
            laws of the \\
         \> sine-Gordon equation.
\end{tabbing}
\end{center}
\subsection{Overview}
The circumstance that the number of conservation laws for the
sine-Gordon equation that were found by the
different methods varies is due to the
varying computational complexity of the determining equations they
generate. The first three methods will find all conservation laws if
memory and time requirements would not matter. The fourth approach
is different in that it generates only necessary conditions which
are often sufficient (if they have the same solution set as the other
methods) but sometimes not. In that case the conditions (\ref{a2})
have additional adjoined symmetries as solutions. It may be that 
only specific linear combinations of them give a conservation law, as
demonstrated with the example in the previous section. The following
comments concentrate on complexity issues and other characteristic
differences between the four approaches.

Arranging the methods as in the table below one can compare rows I,II and
columns A,B.
\begin{center}
\begin{tabular}{|c|c|c|} \hline
 \mbox{\hspace{0.5in}}   & A & B                            \\ \hline
    &   &                                                 \\
 I  & $\mbox{Div}\, P |_{\Delta=0} = 0$ 
    & $ \sum_{\mu,J} (-D)_J \left.
        \left( Q_{\mu} \frac{\partial \Delta_{\mu}}
                            {\partial u^{\nu}_J}
        \right) \right|_{\Delta=0} = 0 \;\;\; \forall \nu $  \\
    &   &                                                 \\ \hline
    &   &                                                 \\
 II & Div $P = \sum_{\nu} Q_{\nu} \Delta_{\nu} $ 
    & $ \sum_J (-D)_J
        \left( \frac{\partial}
                    {\partial u^{\nu}_J} \sum_{\mu}Q_{\mu} \Delta_{\mu}
        \right) = 0\;\;\; \forall \nu $                   \\
    &   &             \\         \hline                           
\end{tabular} \vspace{12pt} \\
Table 5: The four approaches arranged in a table.
\end{center}
{\bf I-II}: The conditions in row I are to be solved in the space of
solutions $(|_{\Delta=0})$, in row II they are not. This means that 
methods of row I can not
be applied if equations or constraints $\Delta_{\mu}=0$ can not be solved for a
leading derivative but methods of row II can. 
Due to these substitutions the conditions in row I have fewer
terms and involve fewer different derivatives of $u$ than 
conditions in row II. The complexity of conditions and the number
of conservation laws up to some order obtained in row I
depend on whether $\Delta_{\mu}=0$ is used to substitute lower order
$u$-derivatives by higher ones or higher ones by lower ones.
There are two reasons for this.

1) Substitutions based on $0 = \Delta$ in $Q$ may give extra restrictions for $Q$.
For example, determining $Q$ for conservation laws of the Korteweg de
Vries equation $0 = \Delta = u_t - u_{xxx} - uu_x$ and restricting $Q$ to be
of $2^{\mbox{\scriptsize nd}}$ order, then a substitution $u_t = u_{xxx} + uu_x$ would imply
$Q = Q(t,x,u,u_x,u_{xx})$, whereas a substitution $u_{xxx} = u_t - uu_x$
would not restrict $Q = Q(t,x,u,u_t,u_x,u_{tt},u_{tx},u_{xx})$.

2) If a lower $u$-derivative is substituted by higher ones using $0=\Delta$
in the conservation law conditions in row I
then such substitutions may increase the order of $u$-derivatives
in which the conservation law conditions have to be satisfied identically.
By that the desired
effect of lowering the number of $u$-derivatives in which the conditions
have to be fulfilled identically is lost. 
For example, condition IB for
$\Delta=u_{tt}-u_{xxt}^{\;\;2}$ is $0=Q_{tt}+2(Qu_{xxt})_{xxt}$ which includes
up to $6^{\mbox{\scriptsize th}}$ order $u$-derivatives (if $Q$ is not of higher
than $3^{\mbox{\scriptsize rd}}$ order). 
By substituting
$u_{tt}=u_{xxt}^{\;\;2}$ the order would increase to seven.

Hence, substituting lower order $u$-derivatives by higher order
$u$-derivatives
gives more over determined conditions for a less general ansatz.
Such conditions are
easier to solve, which may allow higher orders of $Q$ to be
investigated. However, one then may miss conservation laws of 
some order in $P$ or $Q$.

These aspects are not an issue in row II as no substitutions are made there.

\noindent
{\bf A-B}: In column A the single first order conservation law
condition itself is to be solved, and in column B the integrability
conditions of column A, which result when the
conserved current $P$ is eliminated are to be solved. 
Conditions in column B involve
as many equations as there are functions $u^{\mu}$ and they are of the
same order as the highest derivatives of $u^{\mu}$ in
$\Delta_{\nu}=0$. 

The methods in column A compute $P$ and therefore also trivial
conservation laws when determining the general solution of the
determining equations. This may complicate the solution of the
determining equations to some extend. After the determining equations
are solved the trivial conservation laws are easily dropped.
For method IA one simply checks whether Div $P=0$ holds identically
and in method IIA trivial conservation laws have zero integrating
factors $Q_\mu$. The general solution of the determining conditions 
in column B do not generate trivial conservation laws.
Also, conditions in column B are more straightforward
to solve, they can be separated with respect to many high order
$u$-derivatives and yield highly over determined systems. The
disadvantage of methods in column B is that already their
formulation may exceed available computational resources.
Another potential problem with using methods in column B is the
following. If one or more linear PDEs from the over determined
conditions remain unsolved (for example, when investigating the
Burgers equation (\ref{BE1}) then the heat equation remained unsolved
(see equations (\ref{BE1}),(\ref{BE1cl})) then the program will
usually not be able to compute $P^i$ from the $Q_{\mu}$. A way out is to
use methods IB or IIB to get $Q_{\mu}$ and to use that as input to get
$P^i$ from method IIA.

Differences between the approaches are amplified with problems that
involve an increasing number of PDEs and an increasing number of
independent variables. A recommendation for tackling a new single PDE
/ system of PDEs would be:
\begin{itemize}
\item If any ansatz for some or all of the $P^i$ is to be made 
      then the method
      IA has to be used. (An example is the question whether an easily
      integrable conservation law $D_x P^x=0$ with $P^y=0$
      exists.)
\item Otherwise try IIB first.
\item If the conditions become too large to handle then try IB or
      even IIA or IA.
\item If IIB or IB provide the $Q_\mu$ but not the $P^i$ (for example
      because arbitrary functions appear or some conditions remain 
      unsolved) then try IIA using the computed $Q_\mu$ as input.
\end{itemize}

\section{The computer algebra programs} \label{sec9}
The names of computer algebra programs for the four approaches are: 
IA: {\sc ConLaw1},IB: {\sc ConLaw2}, IIA: {\sc ConLaw3}, IIB: {\sc ConLaw4}. 
They and the program {\sc Crack} for solving the over determined
conditions are written in the computer algebra system {\sc Reduce}
({\sc ConLaw1} through {\sc ConLaw4} by the author, {\sc Crack} by
the author and Andreas Brand). From the general solution individual
conservation laws are extracted by picking one arbitrary constant or
function and setting all other arbitrary constants or functions to zero.
The problem is to find, whether all arbitrary constants and functions
are independent or whether some can be dropped without loss of
generality. This problem itself leads  to an over determined system
of conditions which in general is very over determined and easy to
solve. A description of that method is given in \cite{WBM} where also the
computation of $Q$ from $P$ and $P$ from $Q$ is explained.

Compared to other computer algebra programs, the package {\sc Crack}
has a wide variety of techniques for solving
over determined PDE-systems. This allows the following new features 
as compared with other computer programs, a list of which and a 
short description is given in \cite{GH5}:
\begin{itemize}
\item
In all four computer programs $P$ as well as $Q$ is computed.
\item
By solving systems of over determined differential equations
it is possible to find conservation laws with non-polynomial, even non-rational
$P, Q$.
\item
If memory requirements are not too high then the program will make a
definite statement about the existence of conservation laws of a given
order. In the majority of these cases the program will find the
explicit form of the conservation law, otherwise it will return unsolved
equations. 
\item
It is possible to find conservation laws with an explicit 
dependence of $P, Q$ on the independent variables.
\item
There is no limit on the number of DEs nor the number of independent
variables to be investigated for conservation laws other than a limit through the
complexity of computations.
Although not demonstrated in this paper, the program is able to handle
ordinary differential equations (ODEs) as well.
\item
It is possible to determine values of parameters in the DE such that 
conservation laws exist.
\item
For each of the four programs {\sc ConLaw1} through {\sc ConLaw4} 
an ansatz for $P^i$ and/or $Q^{\mu}$ can be input to specify to some
extend conservation laws to be calculated. 
\end{itemize}

A program written by G\"{o}kta\c{s} and Hereman \cite{GH1}
for computing conservation laws of PDEs 
makes a polynomial ansatz for conservation laws and finds the
coefficients in this ansatz by solving a linear algebraic system of
equations. Compared with that, the programs {\sc ConLaw1} through {\sc
ConLaw4} are able to find more general conservation laws and to make a
definitive statement in case the order is not too high to complete
the computations. On the other hand, 
the program of G\"{o}kta\c{s} and Hereman was later
extended to handle differential-difference
systems \cite{GH2},\cite{GH3},\cite{GH4}.

Before showing examples which highlight the
special abilities of {\sc ConLaw1} through {\sc ConLaw4} 
a comment to the treatment of ODEs will be made.
Although all methods and programs are applicable equally well to ODEs, the 
form of the ansatz for the integrating factor $Q$ or for the first integral
$P$ to be made will usually be different.
An $n^{\mbox{\scriptsize th}}$-order 
ODE has always first integrals of order $n-1$ and any 
arbitrary function of first integrals is a first integral as well.
In order to obtain an over determined system of conditions, the ansatz
for a first integral
must not contain functions of all variables $x,u,u',\ldots,d^{n-1}u/dx^{n-1}$
but, for example, a polynomial in $d^{n-1}u/dx^{n-1}$ with arbitrary
functions of $x,u,u',\ldots,d^{n-2}u/dx^{n-2}$ as coefficients or any
other combination of functions of less than $n+1$ variables, see also
\cite{AB4} for more details. Special features of the {\sc ConLaw}
programs that are not available with other programs are demonstrated
with the following examples.

\indent {\it Example}: The It\^{o} equations for two functions
$ u = u(t,x), \;\; v = v(t,x) $ read \cite{Ito}
\begin{eqnarray*}
u_t & = & u_{xxx} + 6uu_x + 2vv_x \\
v_t & = & 2(uv)_x.
\end{eqnarray*}
The first 7 conservation laws calculated by the
program {\sc ConLaw1} which in turn calls
{\sc Crack} to solve condition (\ref{a1}), have the following values
of $P^t$:
\[u, \hspace{5mm} v, \hspace{5mm} u^2+v^2, \hspace{5mm}
u_x^2-2u^3-2uv^2, \hspace{5mm} (4uv^2-v_x^2)/v^3, \]
\[u_{xx}^{\;\,2}-10uu_x^{\,2}-4vv_xu_x+5u^4+6u^2v^2+v^4, \;\;\;
((2vv_{xx}-4uv^2-3v_x^{\,2})^2+16v^6)/v^7.\]
($P^x$ is not shown due to its length. It could be computed easily
from $P^t$.)
What is interesting is that 2 of the 7 conservation laws have a
non-polynomial expression for $P^t$ and as far as the author knows
these conservation laws have not been known so far.

{\it Example}: 
The following equations \cite{KSC}
describe low-frequency Alfv\'{e}n waves
propagating parallel to an external magnetic field
in a relativistic electron-positron plasma \cite{FV}.
Typical for them is the symmetry with respect to interchanging the
two functions $u = u(t,x), v = v(t,x)$ due to the same charge-to-mass ratio
for both kinds of particles.
The equations are
\begin{eqnarray}
\Delta_1 = u_t + r_x = 0,\;\;\;& 
\mbox{with} &\;\;\; r=u(u^2+v^2) + u_{xx}, \nonumber \\
\Delta_2 = v_t + s_x = 0,\;\;\;& 
\mbox{with} &\;\;\; s=v(u^2+v^2) + v_{xx}. \label{KSC}
\end{eqnarray}
The equations themselves have the form of conservation laws. We find
the following additional ones:
\begin{eqnarray*}
4u\Delta_1 + 4v\Delta_2 &\!=\!& D_t[2(u^2+v^2)] + \\
  & & D_x[4uu_{xx}-2u_x^{\,2}+4vv_{xx}-2v_x^{\,2}+3(u^2+v^2)^2] \\
4r\Delta_1 + 4s\Delta_2 &\!=\!& D_t[(u^2+v^2)^2 - 2u_x^{\,2} - 2v_x^{\,2}] + \\
& &  D_x\left[4u_tu_x+4v_tv_x+2u_{xx}^{\;\;2}+2v_{xx}^{\;\;2}+
    4(u^2+v^2)\times \right. \\
& & \;\;\;\;\; \left. \left((3(u^2+v^2)t-x)(uu_t+vv_t)+uu_{xx}+
    vv_{xx}\right)\right] \\
    4(xu-3tr)\Delta_1 &\!+\!& 4(xv-3ts)\Delta_2  = \\
& & D_t\left[3t(\left(2u_x^{\,2}+2v_x^{\,2}-(u^2+v^2)^2\right)+
    2x(u^2+v^2)\right] + \\
& & D_x 2\left[(uu_t+vv_t)\left(-x^2+(u^2+v^2)
       \left(6tx-9t^2(u^2+v^2)\right)\right) \right. \\
& & \;\;\;\;\;\; - 3t(u^2+v^2)^3 + 3x(u^2+v^2)^2 +
           2x(uu_{xx}+vv_{xx}) - 3tr^2 \\
& & \;\;\;\;\;\; \left. - 3ts^2 - 2uu_x - 2vv_x - xu_x^{\,2} - xv_x^{\,2} -
    6tu_tu_x - 6tv_tv_x \right]
\end{eqnarray*}
Whereas the first two are known \cite{FV}, the last one shows an explicit
$x,t$-dependence and is new. Further investigation provides 
that no conservation laws exist with the characteristic 
functions $Q_{\mu}$ of $3^{\mbox{\scriptsize rd}}$ or $4^{\mbox{\scriptsize th}}$ order
(if $u_t, v_t$ are substituted using (\ref{KSC})).

{\it Example}:
The following equation of Gibbons and Tsarev \cite{GT}
\begin{equation}
0 = u_{xx}+u_yu_{xy}-u_xu_{yy}+1
\end{equation}
for $u = u(x,y)$ is unusual in that it has already 5 conservation laws of
first order. The characteristic functions contain $x,y$ explicitly. 
Up to first order they are:
\[1,\;\;\;\; u_y,\;\;\;\; 3u_y^2+2u_x+3x,\;\;\;\; 2u_y^3+3u_xu_y+4u_yx+y, \]
\[10u_y^4+6u_x^2+24u_xu_y^2+20u_xx+30u_y^2x+12u_yy+2u+15x^2,\]
\[3u_y^5+6u_x^2u_y+10u_xu_y^3+18u_xu_yx+4u_xy+
  12u_y^3x+6u_y^2y+12u_yx^2+2u_yu+6xy.\]

{\it Example}: The Liouville equation for a function
$u = u(x,y)$ reads \[ \Delta = u_{xy} - e^u. \]
Conservation laws of order zero found by {\sc ConLaw2} are
\[ (f_x+fu_x) \Delta = D_x(-e^uf) + D_y(f_xu_x + fu_x^{\,2}/2), \;\;\;
  f=f(x) \;\;\; \mbox{arbitrary} \]
\[ (g_y+gu_y) \Delta = D_y(-e^ug) + D_x(g_yu_y + gu_y^{\,2}/2), \;\;\;\,
  g=g(y) \;\;\;\, \mbox{arbitrary}. \]
Because the ansatz made is investigated in full generality, any free functions
in the conservation law will be found if the conditions can be solved completely
by {\sc Crack}. Otherwise
the remaining conditions are returned as in the following example.

{\it Example}: The Burgers equation in the form
\begin{equation}
 \Delta = u_t - u_{xx} - \frac{1}{2}u_x^{\,2} = 0, \;\;\;\; u = u(t,x)
 \label{BE1}
\end{equation} has zeroth order conservation laws
\begin{equation}
 fe^{u/2}\Delta = D_t(2fe^{u/2}) + D_x(e^{u/2}(2f_x-fu_x)) \label{BE1cl}
\end{equation}
with $f = f(t,x)$ satisfying the linear reverse heat equation
$0 = f_t + f_{xx}.$\footnote{Although already used in \cite{AB1}, 
\cite{WBM} this
example is shown again as it also serves to demonstrate an extension
to non-local conservation laws in section \ref{exapp}.}

The occurrence of free functions in the conservation law indicates
linearizability of $\Delta=0$, which is the case for both previous examples.
The following example involves more than 2 variables.

{\it Example}: The Kadomtsev-Petviashvili equation for
$u = u(t,x,y)$ with the abbreviation \[w = u_t + 2uu_x + u_{xxx}\] is
\begin{equation}
0 = \Delta = w_x - u_{yy}. \label{KP}
\end{equation}
Its zeroth order conservation laws include an arbitrary function $c=c(t)$:
\begin{eqnarray}
 c \Delta & = & D_x(cw) + D_y(-cu_y)   \label{KP1}  \\               
 cy \Delta & = & D_x(cyw) + D_y(cu-cyu_y)  \label{KP2}  \\          
 (2cx+c_ty^2) \Delta & = & D_t(-2cu) + \nonumber \\
        & &   D_x\left((2cx+c_ty^2)w-2cu_{xx}-2cu^2\right) +   \label{KP3}  \\
        & &   D_y\left(-(2cx+c_ty^2)u_y+2c_tuy\right) \nonumber   \\
 (6cxy+c_ty^3) \Delta & = & D_t(-6cyu) + \nonumber \\
        & &   D_x\left((6cxy+c_ty^3)w-6cyu_{xx}-6cyu^2\right) +  \label{KP4}  \\
        & &   D_y\left(-(6cxy+c_ty^3)u_y+3c_tuy^2+6cxu\right). \nonumber 
\end{eqnarray}
It is somewhat remarkable that although equation (\ref{KP}) does not involve
$u_t$ but only $u_{xt}$ nevertheless the conserved density $P^t$ in the last
two conservation laws involves $u$ and not $u_x$.

In the following section we give examples for an extension of our
method to compute non-local conservation laws and report on the
possibility to determine parameters in the equation such that
conservation laws exist.
\section{Extending applicability} \label{exapp}
\subsection{Non-local conservation laws}
The implementations of the four methods 
have a common limitation: the characteristic
functions $Q$ and the conserved current $P$ must depend functionally only on a
finite number of derivatives of the $u^\alpha$. No dependencies on
integrals are possible. The same restriction is usually made when
generators of Lie-symmetries are determined for differential equations.
Whereas this restriction is less severe
when calculating symmetries of PDEs, it is
a serious restriction for the determination of conservation laws.
To give an example, Burgers' equation in the form
\begin{equation}
\Delta = u_t - u_{xx} - uu_x = 0, \;\;\;\; u = u(t,x) \label{BE2}
\end{equation}
has as low order conservation law only the trivial one 
$D_tu-D_x(u_x+u^2/2)=0$.
In order to include dependencies on $\int u\,dx$ one could set $u=v_x$
for some function $v(x,t)$ and investigate conservation laws depending on
$v$ and derivatives of $v$. For Burgers' equation such a
substitution alone is not enough. In addition one has to realize that
(\ref{BE2}) can be integrated 
with respect to $x$ to $f(t)_t=v_t-v_{xx}-v_x^{\,2}$
for some function $f=f(t)$. Renaming $v-f \rightarrow u$ gives (\ref{BE1})
and its conservation laws (\ref{BE1cl}).

To give a further example, we consider the
Boussinesq equation describing surface water waves whose horizontal scale is
much larger than the depth of the water \cite{AC},\cite{Hi}
\begin{equation}
u_{tt} - u_{xx} + 3uu_{xx} + 3u_x^{\,2} + \alpha u_{xxxx} = 0. \label{Bou1}
\end{equation}
Calculating conservation laws, using (\ref{Bou1}) to substitute
$u_{xxxx}$, the only characteristic functions
$Q$ up to $4^{\mbox{\scriptsize th}}$ order are $1,x,t,xt.$ 
On the other hand, substituting $u=v_x$,
integrating (\ref{Bou1}) with respect to 
$x$ and renaming $v-f \rightarrow v$ gives
\begin{equation}
v_{tt} - v_{xx} + 3v_xv_{xx} + \alpha v_{xxxx} = 0 \label{Bou2}
\end{equation}
having 2 conservation laws with characteristic functions $1,t$ 
which $x$-differentiated
give the conservation laws above with characteristic functions $1,t.$ In addition
two new conservation laws with characteristic functions $v_x, v_t$ result.
Repeating this step again: $v=w_x$, $x$-integration of (\ref{Bou2}),
$w-f \rightarrow w$ gives
\begin{equation}
w_{tt} - w_{xx} + 3/2 w_{xx}^{\,2} + \alpha w_{xxxx} = 0 \label{Bou3}
\end{equation}
with three third-order conservation laws.
Two of them have characteristic functions $w_{xxx},w_{xxt}$ which 
correspond to
the above conservation laws with characteristic functions 
$v_x,v_t$. In addition one extra conservation law with 
$Q=w_{txx}-w_{txx}w_{xx}+w_{tx}w_{xxx}-\frac{2}{3}w_{ttt}$ exists.

A third example is the  Kadomtsev-Petviashvili equation already discussed
above.\footnote{The hint to try KP for this extension was given by 
Alan Fordy.}
After a substitution $u=v_x$, $x$-integration of (\ref{KP}) and
$v-f \rightarrow v$ the equation is 
\[0 = [v_t+v_{xxx}+v_x^{\,2}]_x-v_{yy}.\] Apart from
conservation laws equivalent to (\ref{KP1}),(\ref{KP2}) three new conservation laws
result with characteristic functions
\[- c_{tt}y^2 - 2c_tx + 4cv_x \]
\[- c_{3t}y^3 - 6c_{tt}xy + 12c_tyv_x + 24cv_y,\]
\[- c_{4t}y^4 - 12c_{3t}xy^2 + 24c_{tt}y^2v_x - 12c_{tt}x^2 + 48c_txv_x
  + 96c_tyv_y + 48c_tv + 144cv_t.\]
Conserved currents are omitted due to their length.
Repeating this transformation again does not yield conservation laws with 
characteristic functions of order less than three.

The purpose of this paragraph was to show that even if computer algebra
programs {\sc ConLaw, Crack} do only allow the investigation of
local conservation laws depending on a finite number of derivatives of the unknown
functions, we still may be able to enlarge the range of search by a
contact transformation and integration of the PDE.

In the next section we extend the computation of conservation laws to
the computation of parameters such that conservation laws exist.
\subsection{Differential equations with parameters}
In applications it is common that the DEs contain parameters and
usually it would be desirable to know conservation laws which are
valid for all possible values of these 
parameters. But as the example below shows, often conservation laws exist
only for special values of parameters. Even if these parameter values are not of 
interest from the application side of view, the conservation laws valid for these
values can at least be used, for example, 
to test numerical code. Another purpose for
determining parameters together with conservation laws could be to
find integrable equations from a more general class of equations. 

The problem to determine parameters such that conservation laws exist
is potentially much harder than determining conservation laws which
are valid for any values of these parameters. This is because the
conservation law determining equations become non-linear. Expressions
may become unmanageably large and many sub cases may have to be
considered. To use {\sc ConLaw1} through {\sc ConLaw4} 
for such calculations one only has to
specify in its call the names of parameters to be computed (more
details in the {\sc ConLaw} manual).

{\it Example}: 
The $5^{\mbox{\scriptsize th}}$-order Korteweg - de Vries equation
\begin{equation}
u_t+\alpha u^2u_x+\beta u_xu_{xx}+\gamma uu_{3x}+u_{5x} = 0    \label{kdv5}
\end{equation}
with constant parameters $\alpha,\beta,\gamma$ includes well known special cases
\cite{FG}, \cite{GH1}, \cite{HIto}, \cite{KW}, \cite{SK}:
for $\alpha=30,\beta=20,\gamma=10$ the Lax equation \cite{Lax},
for $\alpha=5,\beta=5,\gamma=5$ an equation due to Sawata, Kotera \cite{SaKo} and
Dodd and Gibbon \cite{DG},
for $\alpha=20,\beta=25,\gamma=10$ an equation due to 
Kaup \cite{Kaup} and Kupershmidt,
for $\alpha=2,\beta=6,\gamma=3$ an equation due to It\^{o} \cite{Ito}.

The following zeroth and first order conservation laws are calculated
with {\sc ConLaw1} (omitting $P^x$ due to its length in the last two
of these conservation laws): 
\begin{itemize}
\item $Q = 1, \;\; P^t = u$
\item $\alpha=\beta=\gamma=0 : \;\;$ As (\ref{kdv5}) becomes linear, a
conservation law 
is obtained with a characteristic function $Q=Q(x,t)$ satisfying the adjoint PDE
$Q_t + Q_{5x} = 0$ with $P^t = Qu$. 
\item $\alpha=0, \gamma=\beta/3 :\;\; Q = x^2, \;\; P^t = x^2u$ 
\item $\alpha=0, \gamma=\beta/3 :\;\; Q = x, \;\; P^t = xu$ 
\item $\gamma=\beta/2 :\;\; Q = 2u, \;\; P^t = u^2$ 
\item $\alpha = \frac{1}{10}(-2\beta^2+7\beta\gamma-3\gamma^2) : \\
      Q = 60u_{xx}t(\beta-3\gamma) + 6u^2t(2\beta^2-7\beta\gamma+3\gamma^2) + 60x \\
      P^t = 30u_x^2t(-\beta+3\gamma) + u^3t(4\beta^2-14\beta\gamma+6\gamma^2) + 60ux$
\item $\alpha = \frac{1}{10}(-2\beta^2+7\beta\gamma-3\gamma^2) : \\
      Q = 30u_{xx} + 3u^2(2\beta-\gamma),  \\
      P^t = - 15 u_x^2 + u^3(2\beta-\gamma)$
\end{itemize}
We find the same conservation laws as found by the 
program of G\"{o}kta\c{s} and Hereman and in addition
a few conservation laws with explicit $x,t$-dependence.
\section{Summary} \label{summary}
Four approaches to find conservation laws have been compared with
respect to their complexity and other characteristic features. 

In a number of examples, conservation laws have been given, some of them new, which
show that the programs {\sc ConLaw1},...,{\sc ConLaw4} and {\sc Crack}
can be used  
to find local, not necessarily polynomial, conservation laws with explicit variable
dependence and free functions. The programs are, in principle,
applicable to problems with arbitrarily many equations, functions and
variables.

The package is part of the {\sc Reduce} distribution.  The latest
version of the programs including a manual and a test file are
available from \\
{\tt http://lie.math.brocku.ca/twolf/crack/crack.tar.gz} .

\section{Acknowledgement}
The author wants to thank Stephen Anco and especially Alexander Mikhailov 
for many comments. Willy Hereman and George Labahn are thanked for 
their comments on the manuscript. The Symbolic Computation Group at the 
University of Waterloo is thanked for its hospitality during the author's
sabbatical in the fall of 1998.



\section{Appendix: \\ Conservation Laws of the sine-Gordon equation} 
In this appendix conservation laws for the sine-Gordon equation
\[ u_{tx} - \sin(u) = 0 \]
are shown as they have been computed by {\sc ConLaw1-4} and as they are refered to
in tables above. They are not new, we provide them only to illustrate
computer results.
Except for the first conservation law, for all the following there is
an additional conservation law due to the $x \leftrightarrow t$ symmetry.
These results are further examples of the ability of the
programs to compute non-polynomial conservation laws.
\begin{eqnarray}
\!\!\!\!\!\!\!\!\!2(tu_t - xu_x)(u_{tx} - \sin(u)) \!\!&=&\!\! 
  D_t\left[  2\cos(u)t - u_x^2x \right] 
+ D_x\left[- 2\cos(u)x + u_t^2t \right] \label{l1} \\
\!\!\!\!\!\!\!\!\!2u_t (u_{tx} - \sin(u)) \!\!&=&\!\! 
  D_t\left[ 2\cos(u) \right] 
+ D_x\left[ u_t^2  \right]  \label{l2} \\
\!\!\!\!\!\!\!\!\!(8u_{3t} + 4u_t^3 )(u_{tx} - \sin(u)) \!\!&=&\!\! 
  D_t\left[4\cos(u)u_t^2+8u_{tx}u_{tt}-8u_{tt}\sin(u)\right] 
+D_x\left[-4u_{tt}^2+u_t^4\right]\label{l3} 
\end{eqnarray}
%
\begin{eqnarray}
\!\!& &\!\! 2(- 8u_{5t} - 20u_{3t}u_t^2 - 20u_{tt}^2u_t - 3u_{t}^5)
            (u_{tx} - \sin(u)) \label{l4} \\
\!\!&=&\!\! D_t2\left[- 8\cos(u)u_{3t}u_{t} + 4\cos(u)u_{tt}^2 
            - 3\cos(u)u_{t}^4 - 8u_{tx}u_{4t} - 20u_{tx}u_{tt}u_{t}^2
      \right.  \nonumber   \\
\!\!& &\!\! \;\;\;\;\; \left.+ 8u_{4t}\sin(u) + 8u_{3t}u_{ttx} +
      12u_{tt}u_t^2\sin(u)\right]  \nonumber  \\
\!\!&+&\!\! D_x\left[ - 8u_{3t}^2 + 20u_{tt}^2u_t^2 - u_t^6 \right] \nonumber 
\end{eqnarray}
\begin{eqnarray}
\!\!& &\!\! 
8( - 16u_{7t} - 56u_{5t}u_{t}^2 - 224u_{4t}u_{tt}u_{t}
 - 168u_{3t}^2u_{t} - 280u_{3t}u_{tt}^2 - 70u_{3t}u_{t}^4
 - 140u_{tt}^2u_{t}^3 - 5u_{t}^7) \nonumber  \\
\!\!& &\!\!\times (u_{tx} - \sin(u)) \label{l5} \\
\!\!&=&\!\! D_t 8\left[ 
 16u_{6t}\sin(u) - 16u_{tx}u_{6t}
 - 16\cos(u)u_{5t}u_{t} + 16\cos(u)u_{4t}u_{tt}
 - 8\cos(u)u_{3t}^2       \right.  \nonumber   \\
\!\!& &\!\! \;\;\;\;\;  - 40\cos(u)u_{3t}u_{t}^3
 - 20\cos(u)u_{tt}^2u_{t}^2 - 5\cos(u)u_{t}^6 
 - 56u_{tx}u_{4t}u_{t}^2 -  112u_{tx}u_{3t}u_{tt}u_{t}
 \nonumber   \\
\!\!& &\!\! \;\;\;\;\; 
 - 56u_{tx}u_{tt}^3 - 70u_{tx}u_{tt}u_{t}^4
 + 16u_{5t}u_{ttx} - 16u_{4t}u_{3tx}
 + 40u_{4t}u_{t}^2\sin(u) + 56u_{3t}u_{ttx}u_{t}^2
 \nonumber   \\
\!\!& &\!\! \;\;\;\;\; \left.
 + 160u_{3t}u_{tt}u_{t}\sin(u) + 40u_{tt}^3\sin(u)
 + 30u_{tt}u_{t}^4\sin(u)
\right] \nonumber \\
\!\!&+&\!\! D_x \left[
 64u_{4t}^2 - 224u_{3t}^2u_{t}^2 + 112u_{tt}^4
      + 280u_{tt}^2u_{t}^4 - 5u_{t}^8 
\right] \nonumber 
\end{eqnarray}
\begin{eqnarray}
\!\!& &\!\! 
2( - 128u_{9t} - 576u_{7t}u_{t}^2 - 3456u_{6t}u_{tt}u_{t}
 - 7296u_{5t}u_{3t}u_{t} - 6720u_{5t}u_{tt}^2
 - 1008u_{5t}u_{t}^4 \nonumber \\
\!\!& &\!\! 
\;\;\; - 4416u_{4t}^2u_{t}
 - 24192u_{4t}u_{3t}u_{tt} - 8064u_{4t}u_{tt}u_{t}^3
 - 5824u_{3t}^3 - 6048u_{3t}^2u_{t}^3
 - 24864u_{3t}u_{tt}^2u_{t}^2  \nonumber \\
\!\!& &\!\! 
\;\;\; - 840u_{3t}u_{t}^6
 - 6384u_{tt}^4u_{t} - 2520u_{tt}^2u_{t}^5 - 35u_{t}^9) \nonumber \\
\!\!& &\!\!\times (u_{tx} - \sin(u))  \label{l6} \\
\!\!&=&\!\! D_t 2\left[ 
   128u_{8t}\sin(u) 
 - 128\cos(u)u_{7t}u_{t} + 128\cos(u)u_{6t}u_{tt}
 - 128\cos(u)u_{5t}u_{3t} + 64\cos(u)u_{4t}^2
\right. \nonumber \\
\!\!& &\!\! 
\;\;\; 
 - 448\cos(u)u_{5t}u_{t}^3
 - 1344\cos(u)u_{4t}u_{tt}u_{t}^2
 - 1568\cos(u)u_{3t}^2u_{t}^2 - 1344\cos(u)u_{3t}u_{tt}^2u_{t}
\nonumber \\ \!\!& &\!\! \;\;\;  
 - 560\cos(u)u_{3t}u_{t}^5 + 336\cos(u)u_{tt}^4
 - 840\cos(u)u_{tt}^2u_{t}^4 - 35\cos(u)u_{t}^8 - 1008u_{t,x}u_{4t}u_{t}^4
\nonumber \\ \!\!& &\!\! \;\;\;  
 - 576u_{t,x}u_{6t}u_{t}^2 - 2304u_{t,x}u_{5t}u_{tt}u_{t}
 - 4992u_{t,x}u_{4t}u_{3t}u_{t}
 - 4416u_{t,x}u_{4t}u_{tt}^2
 - 128u_{t,x}u_{8t}
\nonumber \\ \!\!& &\!\! \;\;\;  
 - 5184u_{t,x}u_{3t}^2u_{tt}
 - 4032u_{t,x}u_{3t}u_{tt}u_{t}^3
 - 4256u_{t,x}u_{tt}^3u_{t}^2 - 840u_{t,x}u_{tt}u_{t}^6
 + 128u_{7t}u_{tt,x}
\nonumber \\ \!\!& &\!\! \;\;\;  
 - 128u_{6t}u_{3t,x}
 + 448u_{6t}u_{t}^2\sin(u) + 128u_{5t}u_{4t,x}
 + 576u_{5t}u_{tt,x}u_{t}^2 + 2688u_{5t}u_{tt}u_{t}\sin(u)
\nonumber \\ \!\!& &\!\! \;\;\;  
 - 576u_{4t}u_{3t,x}u_{t}^2 + 4480u_{4t}u_{3t}u_{t}\sin(u)
 + 1152u_{4t}u_{tt,x}u_{tt}u_{t}
 + 4032u_{4t}u_{tt}^2\sin(u)
\nonumber \\ \!\!& &\!\! \;\;\;  
 + 560u_{4t}u_{t}^4\sin(u)
 + 1920u_{3t}^2u_{tt,x}u_{t} + 5824u_{3t}^2u_{tt}\sin(u)
 + 3264u_{3t}u_{tt,x}u_{tt}^2
\nonumber \\ \!\!& &\!\! \;\;\;  
\left. + 1008u_{3t}u_{tt,x}u_{t}^4
 + 4480u_{3t}u_{tt}u_{t}^3\sin(u) + 3360u_{tt}^3u_{t}^2\sin(u)
 + 280u_{tt}u_{t}^6\sin(u) \right]  
\nonumber \\
\!\!&+&\!\! D_x \left[
      - 128u_{5t}^2 + 576u_{4t}^2u_{t}^2 - 1280u_{3t}^3u_{t}
      - 3264u_{3t}^2u_{tt}^2 - 1008u_{3t}^2u_{t}^4
      + 2128u_{tt}^4u_{t}^2 
\right. \nonumber \\
\!\!& &\!\! 
\;\;\;\; \left. + 840u_{tt}^2u_{t}^6 - 7u_{t}^{10} \right] \nonumber
\end{eqnarray}

\end{document}